\documentclass[journal]{IEEEtran}
\usepackage{epsfig,color,citesort}
\usepackage{graphicx}
\usepackage{amsmath}
\usepackage{epstopdf}
\usepackage{algorithmic}
\usepackage{algorithm}
\usepackage{amsfonts}
\usepackage{amssymb}
\usepackage{booktabs}
\usepackage{caption}
\usepackage{subcaption}
\usepackage{float}
\usepackage{amsmath}
\usepackage{multirow}
\usepackage[justification=centering]{caption}

\begin{document}

\title{Risk-Aware Resource Allocation for URLLC: Challenges and Strategies with Machine Learning}

\author{Amin~Azari, Mustafa~Ozger, and~Cicek~Cavdar% <-this % stops a space
\thanks{ The authors are with the School
of Electrical Engineering and Computer Science, KTH - The Royal Institute of Technology, Stockholm, Sweden (e-mail: \{aazari, ozger, cavdar\}@kth.se).}% <-this % stops a space
% <-this % stops a space
}
\maketitle

\begin{abstract}
Supporting ultra-reliable low-latency communications (URLLC) is a major challenge of 5G wireless networks. Stringent delay and reliability requirements need to be satisfied for both scheduled and non-scheduled URLLC traffic to enable a diverse set of 5G applications.  Although physical and media access control layer solutions have been investigated to satisfy only scheduled URLLC traffic, there is a lack of study on enabling transmission of non-scheduled URLLC traffic, especially in coexistence with the scheduled URLLC traffic. Machine learning (ML) is an important enabler for such a co-existence scenario due to its ability to exploit spatial/temporal correlation in user behaviors and use of radio resources. Hence, in this paper, we first study the coexistence design challenges, especially the radio resource management (RRM) problem and propose a distributed risk- aware ML solution for RRM. The proposed solution benefits from hybrid orthogonal/non-orthogonal radio resource slicing, and proactively regulates the spectrum needed for satisfying delay/reliability requirement of each URLLC traffic type. A case study is introduced to investigate the potential of the proposed RRM in serving coexisting URLLC traffic types. The results further provide insights on the benefits of leveraging intelligent RRM, e.g. a 75\% increase in data rate with respect to the conservative design  approach  for the scheduled traffic is achieved, while the 99.99\% reliability of both scheduled and non-scheduled traffic types is satisfied.

%Supporting ultra-reliable low-latency communications (URLLC) is a major challenge of 5G wireless networks. Whilst enabling URLLC is essential for realizing many promising 5G applications, the design of communications' solutions for serving such unseen type of traffic with stringent delay and reliability requirements is in its infancy.  In prior studies, physical and media access control layer solutions for assuring the end-to-end delay requirement of scheduled URLLC traffic have been investigated. However, there is lack of study on enabling  transmission of urgent URLLC traffic, especially in coexistence with the scheduled URLLC traffic. This study at first sheds light into the coexistence design challenges, especially the  radio resource management (RRM) problem. It also leverages recent advances in machine learning (ML) to exploit spatial/temporal correlation in user behaviors and use of radio resources, and proposes a distributed risk-aware ML solution for RRM. The proposed solution benefits from hybrid orthogonal/non-orthogonal radio resource slicing, and proactively regulates the spectrum needed for satisfying delay/reliability requirement of each traffic type.
%A case study is introduced to investigate the potential of the proposed RRM in serving coexisting URLLC traffic types.  The results further provide insights on the benefits of leveraging intelligent RRM, e.g. 75\% increase in  data rate for the scheduled traffic is achieved, while the 99.99\% reliability of both scheduled and urgent traffic types is satisfied. 

\end{abstract}

\IEEEpeerreviewmaketitle

\section{Introduction}\label{intro}
\IEEEPARstart{T}{he} fifth generation of wireless networks (5G) has targeted  a  diverse set of wireless services, from enhanced mobile broadband (eMBB) to the Internet of Things (IoT) \cite{1Petar}.  The latter itself could be categorized into two distinct service types, i.e.   massive machine type communications (mMTC), which aims at  connecting everything that benefits from being connected, and  ultra-reliable low-latency communications (URLLC), which requires successful data transmission  within a strictly bounded short time interval \cite{4Risk}.  URRLC is considered as an essential prerequisite of a new wave of services including drone-based delivery, smart factory, remote control, and intelligent transportation systems (ITS) \cite{10Lowlat5G}. 

%-----------------------------------------------------------------
%-----------------------------------------------------------------

%ITS can make a significant impact on society and the daily lives of  human beings by enabling safer and greener transportations.  Towards enabling ITS, providing vehicle-to-everything (V2X) communications  is inevitable. Indeed, V2X communications extend vehicles' field-of-view, which in turn can significantly improve the traffic safety and efficiency, and enable autonomous driving \cite{11DRLV2V}. Autonomous vehicles require fast yet reliable data exchange for coordination among them, platoon management, and overtaking.  In this work, we focus on serving uplink URLLC, especially in the ITS use case.  

Different URLLC applications may have different setup delay tolerances, where  the setup delay is defined as the time from generation of the first URLLC packet at the devices until when the packet is transmitted. Serving URLLC traffic with very-low setup delay tolerance is extremely challenging because this type of traffic cannot wait to be scheduled, e.g. an accident report by a vehicle.    Let us denote this category of  traffic as non-scheduled URLLC traffic. This could be (i) from  connected/disconnected devices which have critical data, usually of short payload size;  (ii) from devices which ask access reservation for subsequent scheduled transmissions of critical data; or (iii) critical device-to-device communications. In the uplink direction, enabling URLLC mandates guaranteeing delay and reliability requirements for both scheduled URLLC transmissions, e.g. continuous control of a drone, as well as the non-scheduled URLLC transmissions, e.g.  immediate accident reports.  The coexistence of scheduled and non-scheduled traffic could happen in many scenarios like vehicular networks, and monitoring and alarm systems \cite{11DRLV2V,10Lowlat5G}.
 Then, the set of  resources allocated to URLLC should be managed to be used for both scheduled and non-scheduled URLLC traffic. Isolation among services via allocating orthogonal slices of resources  is a common practice when dealing with ergodic objectives like throughput. However, in URLLC applications, due to the limitations of time-diversity and crucial need to large spectrum bands for providing frequency diversity, isolation through orthogonal slicing is challenging \cite{10Lowlat5G,3Slicing}.  Recently, serving coexisting  eMBB/URLLC services over limited radio resources has attracted  attentions, and non-orthogonal RRM for serving URLLC  has been proposed for spectrum efficiency \cite{1Petar,2Korea}. While the orthogonal RRM is itself   a complex problem, the non-orthogonal RRM, which requires specifying shared/dedicated resources and scheduling rules over shared ones, is a much more complex problem. Furthermore, to comply with URLLC delay requirements, this problem is needed to be solved in very short time scales. 

Among candidate enablers for solving such  complex, dynamic and time-limited RRM problem, artificial intelligence (AI) is promising \cite{15RRM5G}.  In the light of recent advances in computing and storage technologies, AI is making the leap from traditional pattern recognition use cases  to governing complex systems through advanced machine learning (ML) approaches. 
 While in previous decades, ML,  as a major branch of AI,  has experienced several  up and down periods, this time its  technology readiness level is so high that it has already penetrated to the design  of many complex systems  \cite{15RRM5G}. This motivates us to investigate usefulness of leveraging ML in RRM  for serving the URLLC traffic.  While the traditional ML  consists in a single node which makes decisions in a centralized manner, the delay/reliability requirements of URLLC call for novel, scalable and distributed learning approaches.

\begin{figure}[!th]
	\centering
	\includegraphics[width=3.5in]{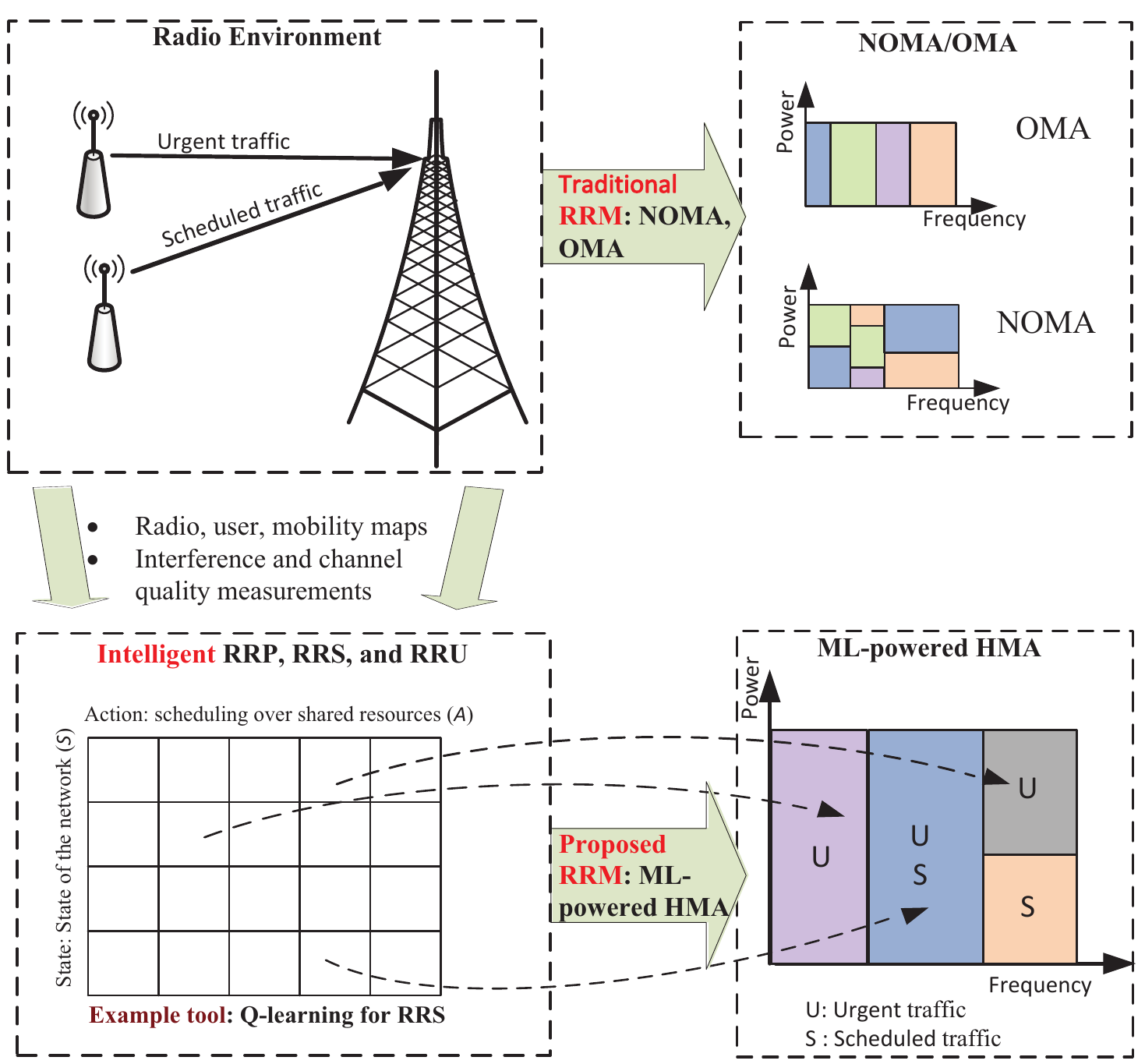} \\	\captionsetup{justification=raggedright,singlelinecheck=false}
	\caption{A graphical illustration of the proposed intelligent RRM solution. The network status, including measurements in the radio environment, is fed to  the intelligent RRM modules, which enable more efficient use of radio resources for scheduled and non-scheduled URLLC traffic types. }
	\label{figfr}
\end{figure} 

\begin{figure*}[ht!]
	\centering
	\begin{subfigure}[t]{0.45\textwidth}
		\centering%{0.2cm 0.01cm 1cm 0.1cm}trim=,clip,
		\includegraphics[width=1\textwidth]{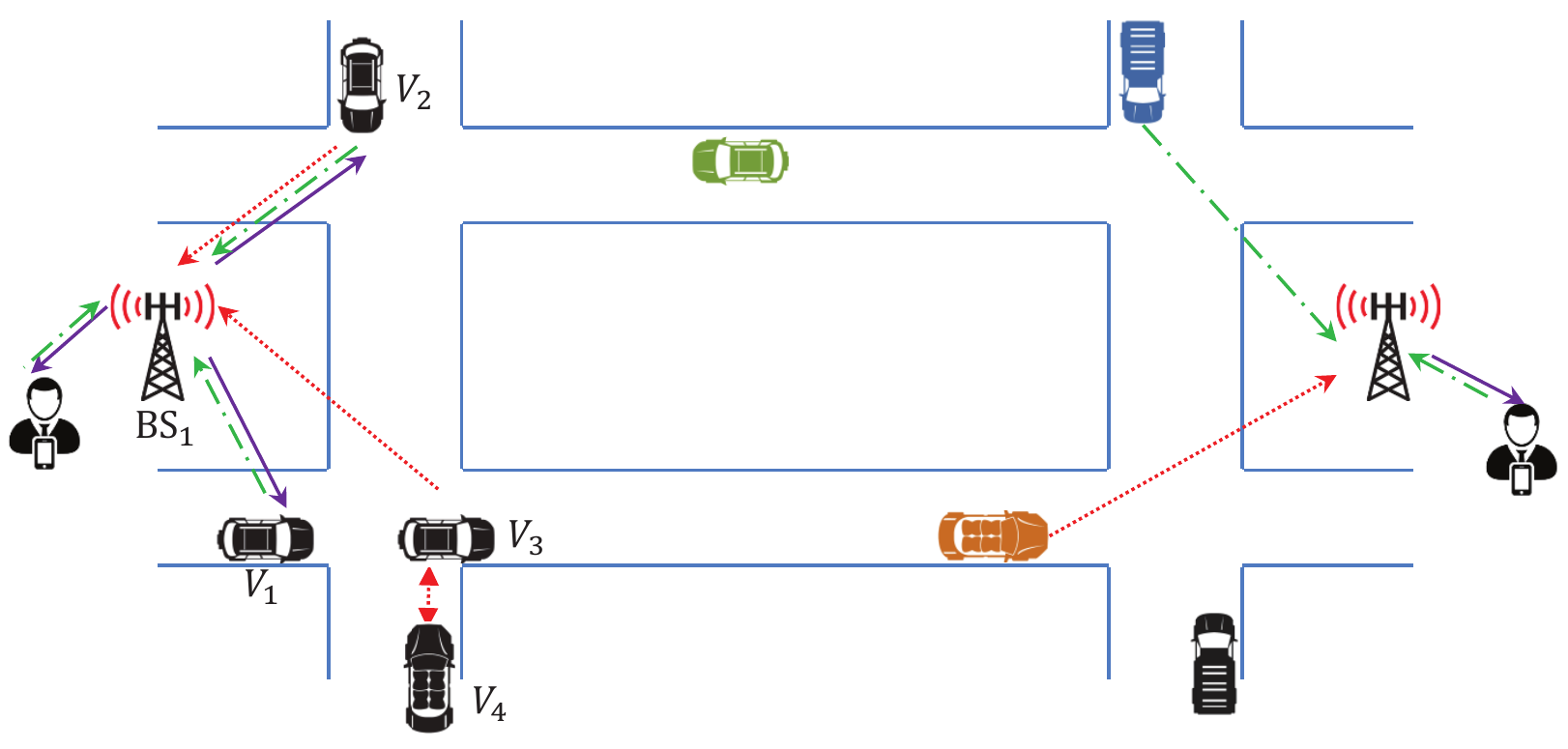} 
		\caption{V2X communications within the ITS use case}\label{fig1}
	\end{subfigure}%
	~
	\begin{subfigure}[t]{0.45\textwidth}
		\centering%{0.2cm 0.01cm 1cm 0.1cm}trim=,clip,
		\includegraphics[width=1\textwidth]{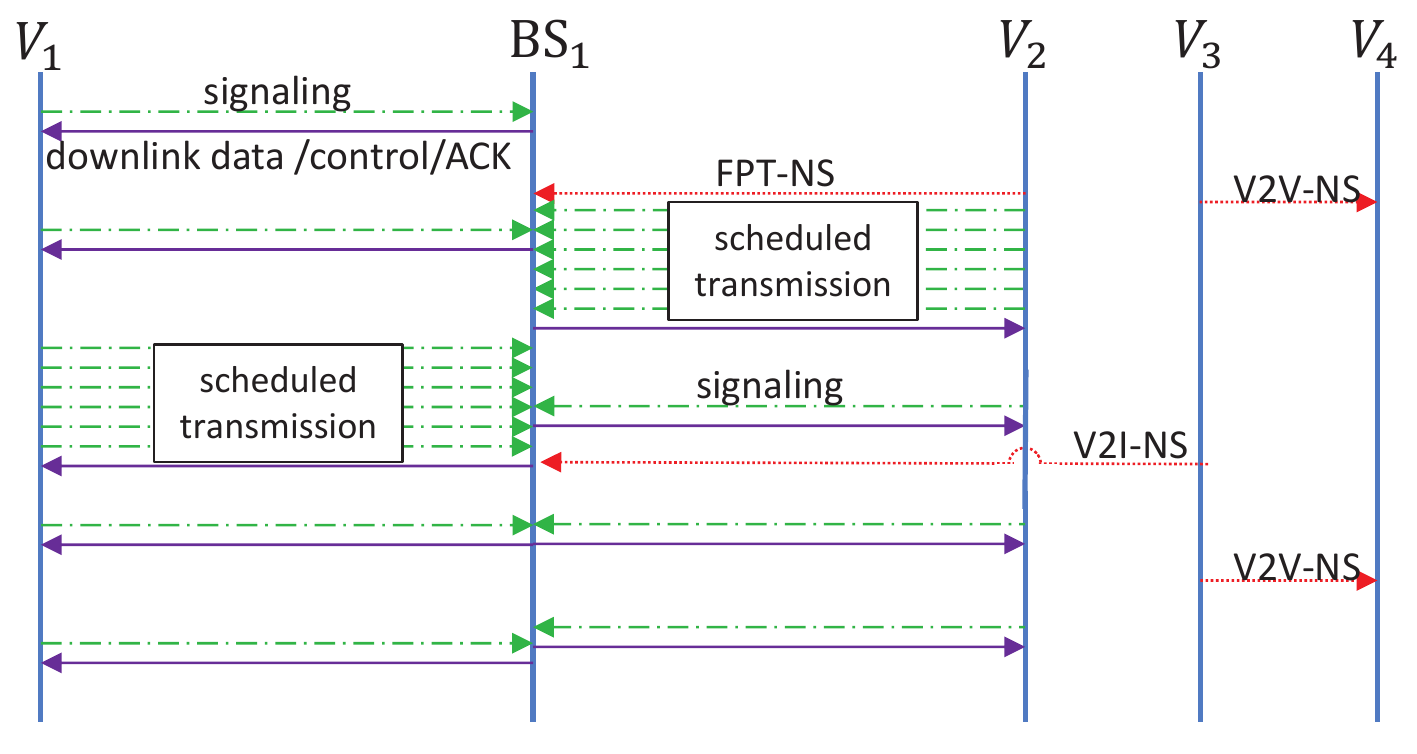} 
		\caption{Communications exchanges }\label{fig2}
	\end{subfigure}%
	\caption{An illustrative example of system model in the ITS use case, including  scheduled (green dashed arrows) and non-scheduled (red dotted arrows) URLLC communications.  }  \label{fig12}
\end{figure*}
% Here, we investigate RRM for serving coexisting  scheduled and non-scheduled URLLC services. 
In this paper, our main focus is on serving urgent  short packet URLLC transmissions along with the scheduled URLLC transmissions. The results derived within this work could be easily extended to the coexistence of URLLC and non-URLLC traffic by modifying the delay/reliability constraints of the scheduled traffic.   The main contributions of this work include:  (i) introduce coexistence of scheduled and non-scheduled URLLC traffic as a challenge to be tackled for realization of URLLC; (ii) introduce a hybrid orthogonal/non-orthogonal multiple access (OMA/NOMA), called HMA scheme, to serve coexisting URLLC traffic; (iii) investigate a distributed learning-powered RRM solution, including intelligent resource provisioning (RRP), scheduling (RRS), and utilizing (RRU) modules, for spectrum efficient yet reliable coexistence management;  and (iv) investigate different sources of uncertainty in proactive RRM, design of a risk-aware ML-powered RRM for  compensating  such uncertainties, and figuring out the amount of radio resources which are sacrificed because of such uncertainties. Fig. \ref{figfr} presents a graphical illustration of the proposed RRM solution and how ML can be mapped to RRM problem and HMA to satisfy URLLC requirements for both scheduled and non-scheduled traffic. We have also present open challenges and potential solutions in RRM for URLLC coexistence.

\section{Enabling URLLC: The Role of Efficient Radio Resource Management}\label{intd}
 In URLLC applications, we usually deal with infrequent bursty traffic, where the arrival time of the burst could not be predicted \cite{1Petar,2Korea}.  Due to stringent delay constraints, once a URLLC packet is generated, the packet must be transmitted immediately without any delay \cite{10Lowlat5G,2Korea}. This abrupt transmission usually having  a short payload size, could be an alarm message, or the first packet transmission (FPT) which will be followed by upcoming scheduled transmissions.  Apart from the non-scheduled  URLLC traffic, we also need serving scheduled URLLC traffic which is not necessary of short payload size \cite{12RRM5G,10Lowlat5G}. The resource allocation to this traffic type should guarantee no packet  drop at the device due to the expiration of data \cite{12RRM5G}.    
 %, i.e. it may include urgent data and reservation for the following traffic
 
Fig. \ref{fig12}(a) represents a graphical illustration of the scheduled and non-scheduled URLLC traffic coexistence in the ITS use-case. In this use case, intelligent vehicles communicate with each other (V2V) as well as with the communication infrastructure (V2I) for traffic efficiency and safety.  One observes that non-scheduled traffic transmissions include (i) FPT of devices which transmit critical data and reserve  access for further messages \cite{10Lowlat5G} (refer to FPT-NS in Fig. \ref{fig12}(b)); (ii) critical message from a device to the access network \cite{2Korea} (refer to V2I-NS in Fig. \ref{fig12}(b)); and (iii) critical V2V communications \cite{11DRLV2V} (refer to V2V-NS in Fig. \ref{fig12}(b)). One further observes that non-scheduled traffic  transmissions are infrequent and could have temporal correlation with the scheduled transmissions (the FPT-NS in Fig. \ref{fig12}(b)). 
%While the above  traffic types and requirements have been introduced in the  ITS use case, they are generic, and  could be also found in other use cases like surveillance and  alarm systems.

Based on the above discussions, enabling URLLC requires tackling three major issues: (i) enabling ultra-reliability in communications, which on the other hand requires diversity; (ii)  enabling low-latency communications, which limits time diversity and enforces frequency diversity; and (iii)  coexistence management of scheduled and non-scheduled traffic, which is due to the fact that frequency resources could not be scaled with the amount of URLLC traffic. The coexistence management calls for  an efficient resource management strategy, without which, scalability of communications systems in serving URLLC service is challenging. 
While in prior studies, the first two issues have been investigated, there is lack of research on the coexistence management within the URLLC services \cite{12RRM5G,2Korea}. Towards serving a coexistence of scheduled and non-
scheduled URLLC traffic, not only the queuing delay violation for scheduled traffic should be minimized, but also the  delay and packet error probability for the first transmission (non-scheduled  traffic) should be reduced dramatically.
Especially, serving non-scheduled traffic  requires revolutionary multiplexing schemes breaking the barrier of interference-free orthogonal transmissions  \cite{1Petar,2Korea}. The need for prompt access may occur in any time instant, e.g. while resources have been reserved for  the scheduled-users, hence, serving these two traffic types together poses coexistence challenges. 

%The main research question is to find an efficient RRM scheme to serve both scheduled and non-scheduled URLLC traffic  over a limited set of  radio resources. 

\subsection{The Hybrid Multiple Access Solution}\label{onh}

The straightforward way of co-serving  scheduled and non-scheduled traffic  is to allocate  them orthogonal sets of radio resources, i.e. OMA. The OMA can be very suboptimal in spectral efficiency because the set of allocated resources to non-scheduled traffic could be unused most of the time due to the bursty nature of such safety-critical messages \cite{1Petar}. On the other hand, insufficient resource allocation to this traffic type results in reliability degradation.  In order to compensate spectral inefficiency of OMA, NOMA  multiplexes different traffic types over a set of shared resources to overcome spectral inefficiency of OMA. However, keeping  isolation among URLLC services with NOMA is challenging due to potential temporal overloading in one service.

Here, we consider a hybrid of OMA/NOMA, namely HMA,  in which,   frequency resources, are divided into three groups: (i) resources allocated to serving the non-scheduled URLLC traffic; (ii) resources allocated to serving the scheduled URLLC traffic; and (iii) resources shared among them. The HMA could be reduced to OMA and NOMA by tuning the amount of shared resources. Over the shared resources, we need to limit the aggregated received interference from the scheduled users.  Furthermore, in order to prevent performance degradation for the scheduled traffic over the shared resources, we need to assure reliability of scheduled data transmission even in presence of non-scheduled  URLLC traffic. Towards this end, proactive or reactive strategies could be used \cite{8PredictiveV2V}. The former aims at making the transmission robust, e.g. by planning for occurring erasure in the channel (erasure coding), while the latter consists in  cancellation of received interference from the non-scheduled  traffic. Regarding the stringent delay requirement of URLLC,  the reactive strategy increases the delay violation probability for the scheduled traffic. Thus,  risk-aware proactive RRM for minimizing reliability degradation is required. 
%\textcolor{red}{In other words, our design target should be shifted from maximizing the average received uplink service to guaranteeing the minimum required uplink service.}
%On the other hand, one must note that in contrast with the legacy average-throughput maximizing objectives used in resource management, here we aim at guaranteeing the performance in worst cases.

\subsection{RRM for the Hybrid Multiple Access}
 To make the HMA solution working, the radio access network (RAN) control center needs to control the traffic load dynamically, and adjust proactively the size of resource pools associated to each traffic stream.  Furthermore, coordination among the BSs is needed to make sure the aggregated intra- and inter-cell interference will not cause an outage. Then, the RRM must be done in a multi-cell level to compensate the inter-cell interference impact on the outage probability.  While a centralized solution seems promising for satisfying the reliability constraint even in cell-edges; the required signaling, including data gathering, forwarding to the center, and decision feedback, can potentially violate the URLLC delay constraints \cite{15RRM5G}. Even in a single cell scenario, due to the potential inter-dependence of the scheduled and  non-scheduled traffic, keeping the outage risk for both traffic types lower than the URLLC requirement is not an easy task.  All in all, regarding the high-mobility of users, their time/location-dependent distributions within the service area, and the bursty nature of  the service requests, the RRM for assuring a minimum service requirement over a wide  service area is highly challenging \cite{12RRM5G,15RRM5G,3Slicing}.

%-----------------------------------------------------------------
%-----------------------------------------------------------------

\begin{figure}[!th]
	\centering
	\includegraphics[width=\columnwidth]{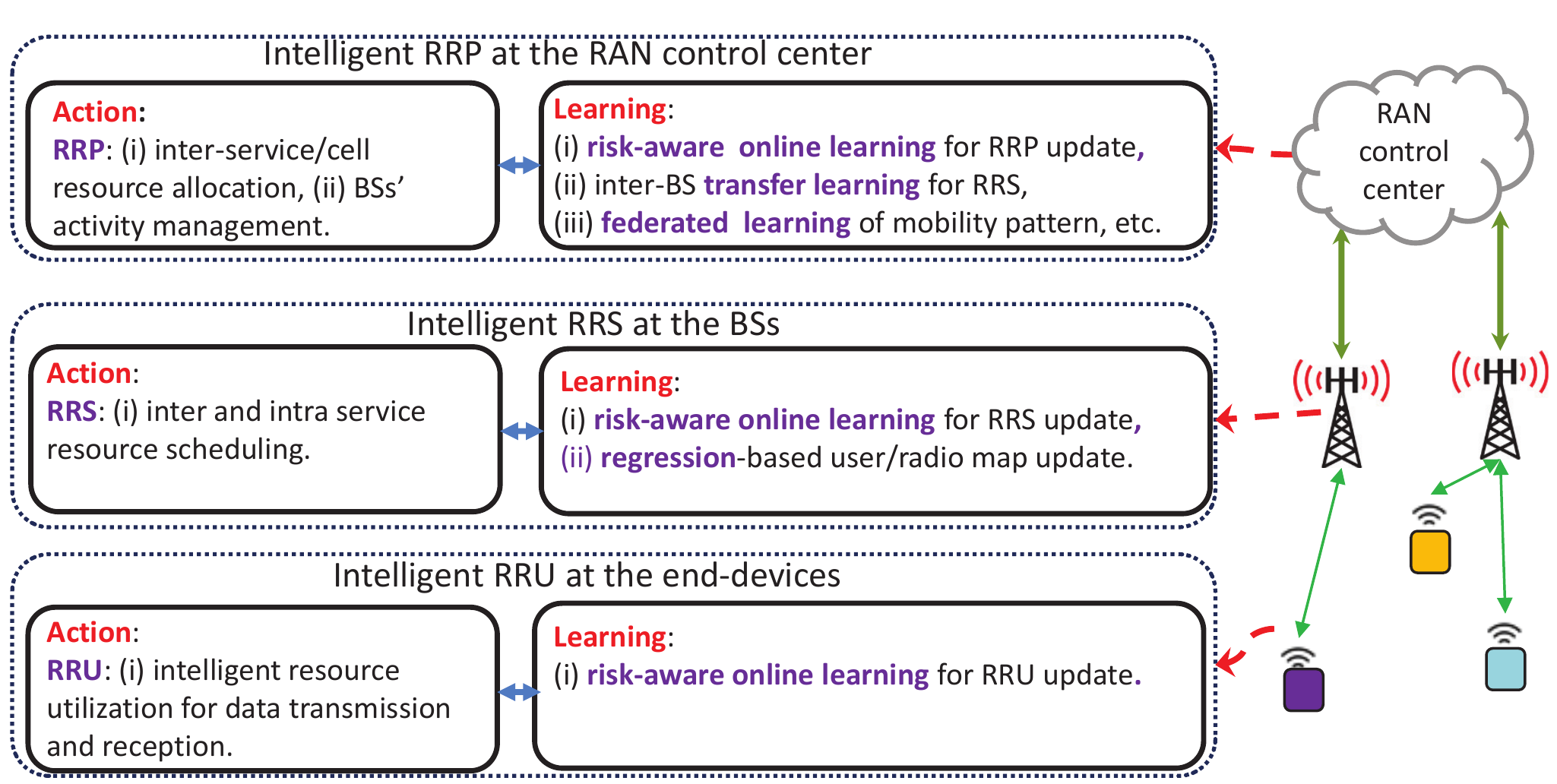} \\	\captionsetup{justification=raggedright,singlelinecheck=false}
	\caption{Distributed ML for radio resource management. The proposed RRM includes (i) RRP at the RAN control center, (ii) RRS at the BSs, and (iii) RRU at the end-devices.}
	\label{fig7}
\end{figure} 

\section{ML in Support of  Efficient RRM }

By broadening the set of services and resources in 5G, control and troubleshooting of  radio access network (RAN) by human resources become too complicated and costly  \cite{15RRM5G}.  Furthermore,  the stringent delay/reliability requirements of newly established URLLC services make them more challenging. Moreover, many decisions taken in serving  users' requests in legacy networks have temporal/spatial correlations, i.e. they are repeated in different locations and time instances. Then, it is wise to leverage the huge volumes of available data in cellular networks, and extract patterns of request arrivals, channel variations of users along mobility patterns, and outage statistics \cite{zoor1,zoor2,15RRM5G}. The derived patterns could  be beneficial in reliable resource provisioning for future service requests. This brings the idea of leveraging ML tools in design and operation control of future wireless networks. 
ML aims at estimating a function, e.g. a rule or pattern, from a set of noisy data  which has been generated by the true function. The recent advances in centralized and distributed computing,  storage, and learning algorithms have improved the position of ML as a design and optimization tool significantly \cite{15RRM5G}.

\subsection*{The Proposed ML-Powered RRM}
To manage network resources for serving scheduled and non-scheduled  URLLC traffic, one needs to identify  the set of dedicated resources to non-scheduled  traffic,  the set of shared resources, and  the scheduling policy over the shared and dedicated resources. 
Once the set of resources and scheduling rules are determined, the  scheduler  allocates  resources to users based on their rankings. The rank of each user  depends on its delay budget and  the expected level of received interference from the user over the shared resources. We propose to break the intelligent RRM problem into 3 sub-problems: radio resource provisioning (RRP), radio resource scheduling (RRS), and   radio resource utilizing (RRU). 

Fig. \ref{fig7}  represents a graphical representation of the proposed hierarchical learning solution. In the proposed solution, a proactive resource provisioning based on  training and prediction is done at the RAN control center. RRP at the   RAN control center benefits from (i) {\textit{transfer learning}}, i.e. making decision for a service area based on  decisions taken in neighboring areas or previous time instances \cite{5Transfer}; (ii) {\textit{federated learning}}, i.e. leveraging the huge volumes of available records of requests/responses  at the edges for extracting common rules and policy sets \cite{15RRM5G}; and (iii) {\textit{risk-aware learning}} from past RRP decisions for outage minimization of coexisting services and traffic types. Distributing the learning task between BSs and the RAN control center  aims at compensating partial observability of BSs in cellular networks. As BSs have access to local information, i.e. they observe only a limited part of the service area, and URLLC users could be highly mobile, resource provisioning at the BS-level could be less effective. For example, a BS cannot predict the arriving of a vehicle platoon which is currently in the neighboring service area. Then, distributed learning can further enable transfer knowledge of arrival of such demanding users, and sharing of respective effective resource scheduling policies. As an example, such learning may result in the knowledge for the RAN control center to activate some dormant BSs to balance the scheduled and non-scheduled traffic serving responsibilities among BSs, and hence, to minimize the probability of outage. 

Furthermore, in the proposed intelligent RRM solution, BSs decide how to control access of the scheduled users over the shared and dedicated resources to minimize the queuing delay violation for them, while complying with the maximum allowed interference constraint over the shared resources.  This distributed control  allows BSs to react to instant changes in the channel quality of connected users, and schedule them based on their remaining delay budgets. 
Finally, the intelligent usage of radio resources for reliable data transmission/reception is done at the device side. Devices can decide how to allocate their power resources to the set of shared and dedicated frequency resources to ensure the highest possible reliability. This is especially the case in non-scheduled URLLC transmissions, either to the BS or to another device/vehicle, where devices can benefit from reinforcement learning  for improving reliability of their transmissions \cite{SOIoT}.

The proposed RRM  requires a set of ML solutions at the RAN control center, BS, and device-side for addressing  the challenges posed in serving the URLLC traffic, as described in Section \ref{intro}. In the following section, we describe some of these solutions, as well as the open problems. These solutions have been also depicted in the learning boxes of  Fig. \ref{fig7}.

\section{ML Solutions Embedded in the Proposed RRM}
\label{prop}
\subsection{Federated Learning of User and Radio Maps for RRP}
The RRP for URLLC traffic at the RAN control center requires knowledge about the outage risk of potential users. When we consider no a priori information about the users, the design is based on the worst case scenario, and hence, huge volumes of frequency resources are required to satisfy a level of QoS. Here, we investigate the combined use of radio and user maps in achieving an estimate of the outage risk of URLLC users.  
%09159607670 Hoseini--> 
%09156673653 Khazaei--> 8.5

\subsubsection{Radio Map}  They are expected to be an essential 5G component for  enabling agile yet efficient resource allocation for mobile use cases like autonomous driving \cite{6Radiomap}. For a given service area, radio map includes  information like radio signal strength, delay spread, and coexisting interference  level.  Regarding  the  inevitable changes in the environment,  radio maps need to be updated once a time.    In the proposed hierarchical RRM solution, this is the role of the RAN control center to coordinate continuous update of the radio map, and to manage location-based inter-update time intervals. Furthermore, there might  be time periods during which, no update of signal strength  from a specific region is sent to the BSs. This could be due to temporal absence of users in a location for an extended period of time. In such cases, this is the role of the involving BSs to predict and update the received signal strength in that location, e.g. by leveraging  temporal/spatial regression as described in \cite{6Radiomap}.

\subsubsection{User Map} With the ever increasing number of connected devices and the volume of online data, the locations and mobility patterns of many vehicles, specially public transportation systems, are known or could be learnt \cite{7Location5G}. Further than offline learning of mobility patterns, the RAN control center can coordinate information exchange among  BSs in a service area  to let them collaboratively construct and update a common user map.    Having a map of users and mobility patterns, enables prediction of presence of users in different regions of the service area for an extended period of time \cite{7Location5G}. The map of users requires continuous update to be effective in URLLC applications. In our proposed structure, this is the role of the RAN control center to keep the consistency of the global user map, and coordinate the inter-update time intervals among the involving BSs.

%\subsubsection{User/Radio Map for  RRP}
 Once the position/speed/direction information of a user is known, and the radio map of the respective environment is given, we are able to estimate path-loss and shadowing components of its wireless communication channel. Then, the uncertainty of the channel scales down to the small-scale fading. Thus, we will be able to predict the received power at the BS as a function of transmit power of users. Within the context of URLLC, this type of information is useful for early identification of a user  which is expected to enter a crowded/low-coverage area. Thus, the RAN would be able to configure the network -manage the radio/energy resources- to lower the risk of communication outage for the respective users.  The radio maps will further enable user management based on regional characteristics.  This means that once devices enter a specific region, they could be granted higher-level support, e.g. dual connectivity, or extra resources for their transmissions could be reserved, or their urgent transmissions could be handed over to multiple neighboring BSs. 
%This ensures finding policies that not only achieve a high  return, but also the return is distributed closely to its means value.
\subsection{Risk-Aware Learning for RRS}

Introduction of URLLC into cellular networks mandates a transition from average-utility network design into risk-aware network design, where a rare event may incur a huge loss \cite{4Risk}. Instead of legacy ML algorithms, which aim at maximizing the average return of agents, or equivalently minimizing the average regret of agents, we aim at minimizing the risk of huge loss. The loss could be defined in different ways as a function of the return. When serving solely one traffic type, the risk-aware learning-powered scheduling enforces maximizing the first moment of the return, i.e. reliability, while minimizing its higher moments, e.g. the variance.  Minimizing higher moments of the return, i.e. optimizing the worst case return,  could be achieved by defining the utility function as a non-linear function of  the return, $R$. A well-known example is  exponential utility function in which, instead of maximizing the expected value of $R$,  the objective is to maximize the expected value of $\exp(\text{-}R)$ \cite{4Risk}. In our proposed RRM solution, scheduling of shared radio resources is done by leveraging risk-aware learning at the BSs. In other words, the scheduling task is performed such that the risk of outage for scheduled and non-scheduled traffic remains meets  the  URLLC's  requirements.

\begin{figure*}[t!]
	\centering
	\begin{subfigure}[t]{0.49\textwidth}
		\centering%{0.2cm 0.01cm 1cm 0.1cm}trim=,clip,
		\includegraphics[trim={0.5cm 0.01cm 1.1cm 0.15cm},clip,width=1\textwidth]{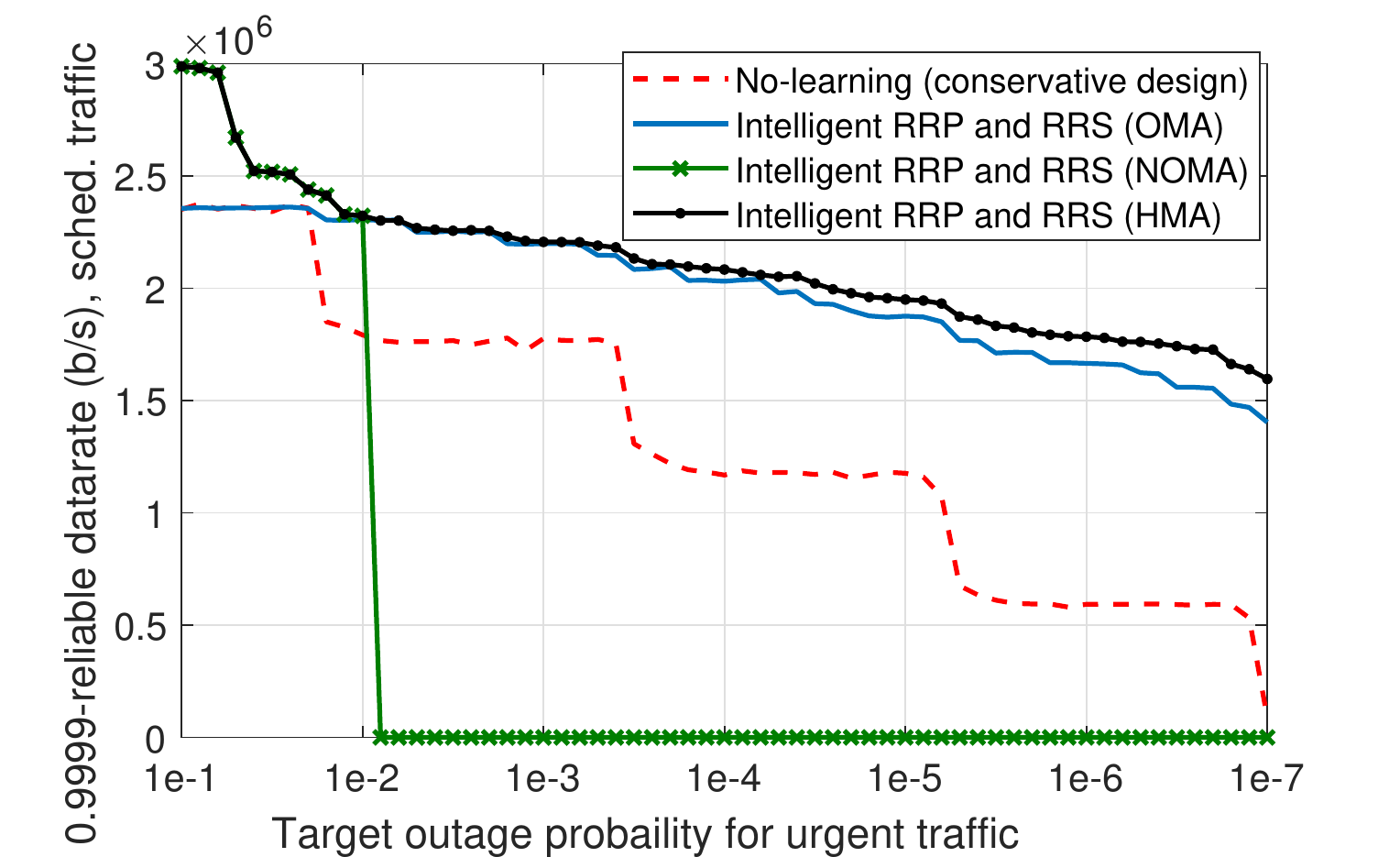} 
		\caption{Intelligent RRP and RRS: reliable data rate  for the scheduled traffic  versus the  outage probability for the non-scheduled traffic (single-shot transmissions). Increase in the required reliability  highlights the merits of intelligent RRM and HMA.}\label{fig5n}
	\end{subfigure}%
	~
	\begin{subfigure}[t]{0.49\textwidth}
		\centering%{0.2cm 0.01cm 1cm 0.1cm}trim=,clip,
		\includegraphics[width=1\textwidth]{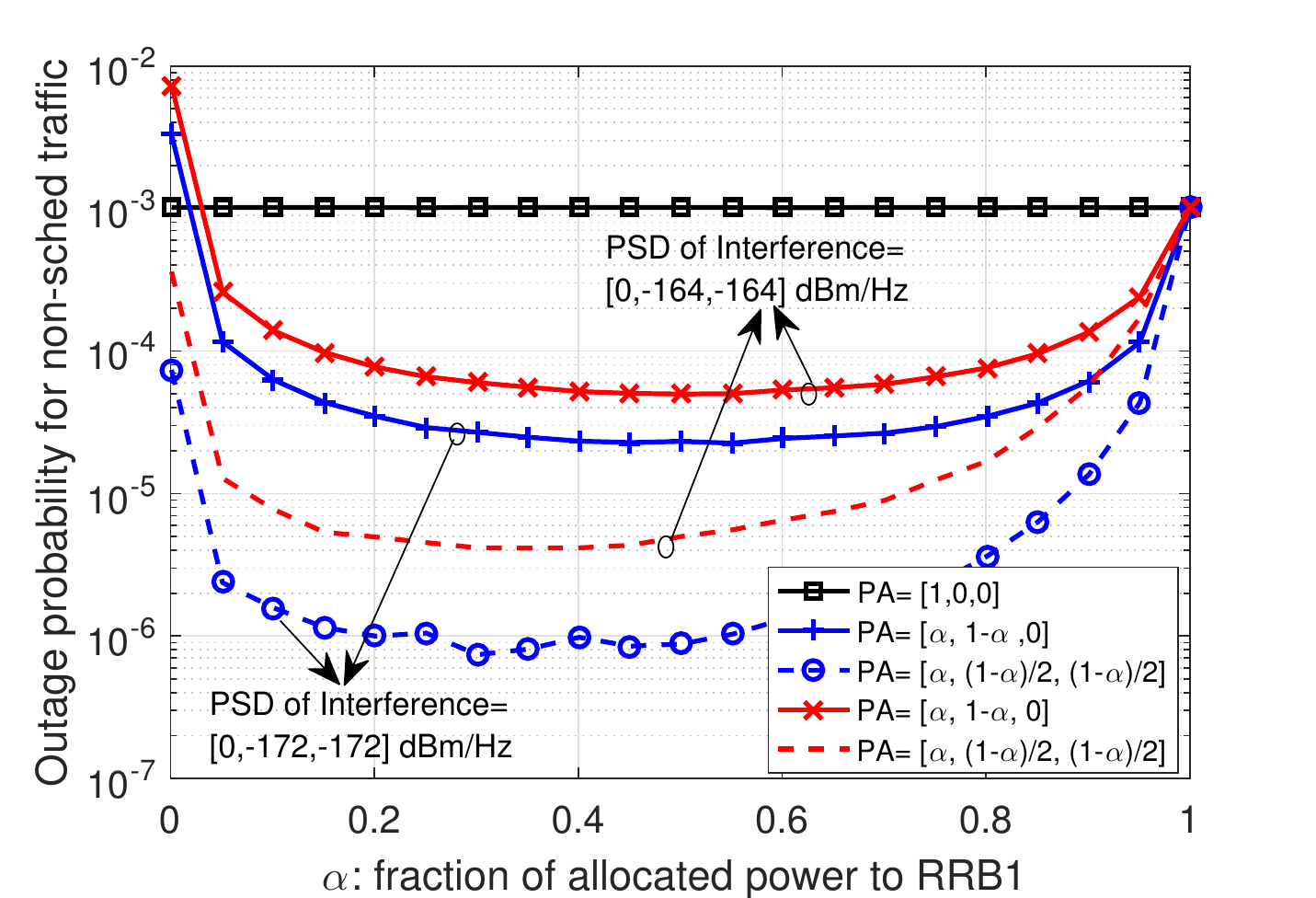} 
		\caption{Intelligent RRU: outage probability for different power allocation (PA) decisions. PA=$[\alpha,x,y]$ represents a  decision in which, $\alpha$ fraction of power is used for the dedicated RRB, and $x$ and $y$ fraction is used for the other two shared RRBs.}\label{fig3n}
	\end{subfigure}%
	\caption{Performance evaluation of the proposed   RRM.}  \label{fign}
\end{figure*}

\subsection{Open Challenges and Potential Solutions}

The main challenge in utilizing intelligent RRM consists in large dimensionality and complexity. The heterogeneity of services, QoS requirements, and diverse set of available resources make the RRM a very complex problem. This complexity has a significant impact on the convergence time of the learning algorithms. Furthermore,  the high mobility of users demanding URLLC service, especially in the ITS use case, and the partial observability of the network by each BS, put further constraints to the RRM problem. 
%Regarding the delay and complexity constraints, the role of on-device and edge learning for decision making must be highlighted. 
In order to  overcome the shortcomings of ML in timely finding  optimal decisions, it is important to compare it against the way human beings learn. Human beings benefit from planning and reasoning for reducing the search space and explore the solution in most probable areas. In  contrast, ML approaches treat  the search space unbiased, and in the exploration phase, select all actions randomly to get an estimate of their respective payoffs. One way to reduce the search space, and speed up the learning process, is to add smart planning to the learning algorithms. Towards this end, one may benefit from the human-resources, and integrate them  in the learning loop, to reduce the search space, and prevent BSs in taking risky/time-consuming actions during the exploration process. 
Another major challenge in utilizing intelligent RRM consists in exchanging information among different network entities, e.g. from a BS to the RAN control center, which is energy and radio resource consuming. Towards addressing this problem, designing communication-efficient learning algorithms, e.g. federated learning in which only the learnt models are communicated \cite{15RRM5G}, are of crucial importance.

 \begin{table}[b!]
 \centering \caption{Parameters of the case study.}\label{tb1}
\begin{tabular}{p{2 cm}p{5.8 cm}}\\
\toprule[0.5mm]
\bf{Parameters} & \bf{Values}\\
\midrule[0.3mm]
Service area& Circular area, radius 3 Km, BS at the center\\
Traffic& Scheduled: saturated; Non-scheduled: exp(0.01)\\
Radio resources& 5 radio resource blocks (RBBs), each 180 KHz\\
Transmit power& Scheduled: 21 dBm, Non-scheduled: 23 dBm\\
Intelligent RRM & Q-learning: learning rate=0.1, exploration rate=0.85, $Q$($S$,$A$): value of  state-action combinations\\
Q-learning for RRP and RRS&  $S$: state of the network including reliability requirements and position of users, $A$: scheduling of a  user  over shared resources\\
Q-learning for RRU&  $S$: state of the network including interference level on RRBs, $A$: power allocation over RRBs\\

\bottomrule[0.5mm]
\end{tabular}
\end{table}

\section{ML-Powered  RRM: A Case Study}
In this section, we demonstrate how to take advantage of intelligent RRM  for spectrum-efficient yet reliable serving of scheduled and non-scheduled URLLC traffic.   
First, we investigate performance of intelligent RRP and RRS  (Fig. \ref{fign}(a)). Consider a single-cell scenario in which URLLC users generate and transmit  non-scheduled URLLC traffic (infrequent) as well as  scheduled URLLC traffic. The minimum and maximum experienced  path-loss in the service area  are denoted by $\text{-}70$ and $\text{-}120$ dB, respectively. Other simulation parameters could be found in Table \ref{tb1}. Based on the outage risk of the most critical device with potential non-scheduled URLLC traffic, the required outage probability for the non-scheduled traffic, and the learnt risk of outage in the service area, the ML-powered RRP and RRS modules are  implemented and used for resource management, as described in Section \ref{prop} and Table \ref{tb1}.  Fig. \ref{fign}(a) represents the level of reliable data-rate achieved for the scheduled traffic, versus the required outage probability for the non-scheduled traffic. The dashed-curve represents a conservative RRM solution in which, resource allocation is done based on the  risk to the cell-edge user.  For easier interpretation of the results, the decoupled gains of RRM with OMA, NOMA,  and HMA have been  presented. OMA represents orthogonal allocation of RRBs to scheduled/non-scheduled traffic based on the online outage risk-level, in contrast with the conservative design (the dashed curve). NOMA represents the case in which, all RRBs could be reused by the scheduled traffic, and hence, there is no dedicated RRBs for the non-scheduled traffic.  HMA represents  the advanced form of NOMA in which, once reliability requirement of the non-scheduled traffic increases, it dedicates some resources to the non-scheduled traffic. To quantify the achieved gain by intelligent RRM, let us focus on the 99.99\% reliability requirement for both scheduled and non-scheduled  traffic types. One observes that 75\% increase in reliable data rate for the scheduled traffic is achieved by intelligent RRM, while the  reliability of both traffic types is satisfied.  One further observes, the higher the outage risk, the higher room for spectrum efficiency by intelligent RRM. Also, we see that ultra-reliable serving of the non-scheduled traffic mandates dedicating some RRBs, and beyond the target reliability level of 99\%, the NOMA scheme fails in serving the non-scheduled  traffic.  Furthermore,  the benefits of HMA increase by increase in the required reliability level. Finally, we observe that uncertainty in the wireless channel, as well as user demand, mandates reserving more radio resources for guaranteeing reliability performance, which on the other hand reduces the overall spectrum efficiency. 

%Once the set of shared resources is known, the radio resource scheduler leverages the  received scheduling policy from the RAN simulator,  assigns scheduled traffic over shared resources, and updates the scheduling policies based on the observed risk to the both traffic types.
Now, let us examine the benefit of intelligent radio resource utilization. We assume a typical device with non-scheduled URLLC traffic experiences a distance-dependent path-loss of $\text{-}80$ dB, and needs to combat small-scale fading as well as path-loss for achieving reliable communications. Out of $5$ RRBs, $3$ could be used for urgent transmission of this device, where the power spectral density (PSD) of scheduled traffic's interference over them is denoted by $[0, \text{-}172,\text{-}172]$ dBm/Hz. In other words, the first one is dedicated and the other two are shared with the scheduled traffic. Fig. \ref{fign}(b) represents the outage probability versus the fraction of allocated power to the dedicated RRB, i.e. $\alpha$. In this figure, PA=$[\alpha,x,y]$ represents the power allocation decision over the RRBs by HMA, which reduces to OMA and NOMA when PA= [1,0,0] and PA=[1/3,1/3,1/3], respectively. One sees that the best action is to distribute power equally on all RRBs. This is because the PSD of interference is comparable with the PSD of noise, i.e. $\text{-}174$ dBm/Hz.  On the other hand, one observe that when the PSD of interference is $[0, \text{-}164,\text{-}164]$ dBm/Hz, i.e. interference is much stronger than the noise, the best action is to use half of the power over the dedicated RRB, and divide the other half among the two shared RRBs. In both cases, one observes that an {\textit{intelligent}} selection of $\alpha$, could bring a huge gain for HMA in comparison with the OMA and NOMA.

%-----------------------------------------------------------------
%-----------------------------------------------------------------	 
\section{Conclusion}
Regarding the stringent reliability and delay requirements of URLLC, there is still lack of communications solutions enabling the URLLC service. One major challenge consists in addressing URLLC’s scalability and co-existence with other URLLC/non-URLLC services. Here, we have investigated coexistence of scheduled and  non-scheduled URLLC services and discussed challenges to be addressed for satisfying their stringent requirements in co-existence scenarios. Then, we have proposed a hybrid multiple access (HMA) solution for addressing such issues in a spectral/energy efficient way. In the heart of our proposed solution, we have further leveraged a distributed hierarchical ML approach for proactive RRM to different URLLC traffic streams. Finally, we have provided a case study to demonstrate the potential of leveraging ML in serving coexisting URLLC traffic over limited radio resources. The results indicated that even for target outage probability of $10^{-7}$ for non-scheduled URLLC traffic, approximately more than $6$ times data rate is achieved in comparison to conservative design for scheduled URLLC traffic with $99.99 \%$ reliability thanks to our ML-powered HMA solution. 

%Regarding the stringent reliability and delay requirements of URLLC,  there is still lack of communications solutions  enabling the URLLC service. One major challenge consists in addressing URLLC's  scalability and coexistence with other URLLC/non-URLLC services. Here, we have investigated coexistence of urgent non-scheduled and scheduled URLLC services, and  discussed challenges to be addressed for satisfying their stringent requirements. Then, we have proposed a hybrid multiple access solution for addressing such issues in a spectral/energy efficient way. In the heart of our proposed solution,  we have further leveraged a distributed hierarchical ML approach for proactive RRM to different URLLC traffic streams. Finally, we have provided a case study to demonstrate  the potential of leveraging ML in serving coexisting URLLC traffic over limited radio resources. The results indicated the gains that could be achieved by utilizing the proposed RRP, RRS, and RRU modules  in comparison with the conservative non-learning  operation. 

\section*{Acknowledgment}
This work is supported in part by the Celtic Plus Project SooGreen (Service Oriented Optimization of Green Mobile Networks).

\ifCLASSOPTIONcaptionsoff
  \newpage
\fi

\begin{IEEEbiographynophoto}
	{Amin Azari} received the B.Sc. and M.Sc. degrees in electrical and communication systems engineering from University of Tehran, Iran in 2011 and 2013, respectively, and the Ph.D. degree in information and communication technology from KTH Royal Institute of Technology, Sweden in 2018. Currently, he is a postdoctoral researcher at Stockholm University. His research interests include 5G radio access network design, Internet of Things and machine learning.
	
\end{IEEEbiographynophoto}

\begin{IEEEbiographynophoto}
{Mustafa Ozger} received his B.Sc. degree in electrical and electronics engineering from Middle East Technical University, Ankara, Turkey, in 2011, and his M.Sc. and Ph.D. degrees in electrical and electronics engineering from Koc University, Istanbul, Turkey, in 2013 and 2017, respectively. Currently, he is a postdoctoral researcher at KTH Royal Institute of Technology. His research interests include wireless communications and the Internet of Things.
\end{IEEEbiographynophoto}

\begin{IEEEbiographynophoto} 
{Cicek Cavdar}is an assistant professor at the School of EECS at KTH in Sweden. She has been leading the “Intelligent Network Systems” research group under the Radio Systems Lab focusing on design and planning of intelligent network architectures, direct air-to-ground communications, and IoT connectivity platforms. She has been coordinating the EU EIT Digital project “Seamless DA2GC in Europe,” which has resulted in successful technology transfer cases to industry. She served as Symposium Chair for IEEE ICC GCSN 2017.
\end{IEEEbiographynophoto}

\end{document}